\documentclass[conference]{IEEEtran}
\IEEEoverridecommandlockouts
\usepackage{cite}
\usepackage[T1]{fontenc}
\usepackage[utf8]{inputenc}
\usepackage{amsmath,amssymb,amsfonts}
\usepackage{algorithmic}
\usepackage{algorithm}
\usepackage{graphicx}
\usepackage{caption}
\usepackage{subcaption}
\usepackage{tabularx}
\usepackage{textcomp}
\usepackage{xcolor}
\usepackage{dsfont}
\usepackage{nopageno}
\usepackage{url}
\def\BibTeX{{\rm B\kern-.05em{\sc i\kern-.025em b}\kern-.08em
    T\kern-.1667em\lower.7ex\hbox{E}\kern-.125emX}}
\begin{document}

\newtheorem{theorem}{Theorem}
\newtheorem{lemma}{Lemma}
\newenvironment{proof}{\paragraph{Proof:}}{\hfill$\square$}

\let \ALP \mathcal 
\let \VEC \boldsymbol
\renewcommand{\Re}{\mathbb{R}}
\newcommand{\transpose}{\mathsf{T}}
\newcommand{\ind}[1]{\mathds{1}_{#1}}
\newcommand{\beq}[1]{\begin{align} #1 \end{align}}
\newcommand{\beqq}[1]{\begin{align*} #1 \end{align*}}
\renewcommand{\Re}{\mathbb{R}}
\newcommand{\Na}{\mathbb{N}}
\newcommand{\Z}{\mathbb{Z}_{+}}
\newcommand{\ex}[1]{\mathds{E}\left[#1\right]}
\newcommand{\pr}[1]{\mathds{P}\left\{#1\right\}}
\renewcommand{\sp}{\texttt{span}}
\newcommand{\pas}{\;\;\mathbb{P}-\text{a.s.}}
\renewcommand{\forall}{\text{ for all }}

\newcommand{\ag}[1]{{\color{red} \textbf{Ab:} {#1}}}
\newcommand{\sbv}[1]{{\color{blue} \textbf{Sh:} {#1}}}
\newcommand{\av}[1]{{\color{green} \textbf{Ar:} {#1}}}

\newcommand{\ourAlgorithm}{Cobalt}

\title{Cobalt: Optimizing Mining Rewards in Proof-of-Work Network Games \\

}

\author{\IEEEauthorblockN{Arti Vedula, Abhishek Gupta and Shaileshh Bojja Venkatakrishnan}
\IEEEauthorblockA{
\textit{The Ohio State University}\\
\{vedula.9, gupta.706, bojjavenkatakrishnan.2\}@osu.edu}
}
\IEEEpubid{\makebox[\columnwidth]{978-8-3503-1019-1/23/\$31.00~\copyright2023 IEEE \hfill} \hspace{\columnsep}\makebox[\columnwidth]{ }}

\maketitle
\IEEEpubidadjcol
\pagestyle{plain}
\begin{abstract}
Mining in proof-of-work blockchains has become an expensive affair requiring specialized hardware capable of executing several megahashes per second at huge electricity costs.  
Miners earn a reward each time they mine a block within the longest chain, which helps offset their mining costs. 
It is therefore of interest to miners to maximize the number of mined blocks in the blockchain and increase revenue.   
A key factor affecting mining rewards earned is the connectivity between miners in the peer-to-peer network.  
To maximize rewards a miner must choose its network connections carefully, ensuring existence of paths to other miners that are on average of a lower latency compared to paths between other miners. 
We formulate the problem of deciding whom to connect to for miners as a combinatorial bandit problem. 
Each node picks its neighbors strategically to minimize the latency to reach 90\% of the hash power of the network relative to the 90-th percentile latency from other nodes.
A key contribution of our work is the use of a network coordinates based model for learning the network structure within the bandit algorithm. 
Experimentally we show our proposed algorithm outperforming or matching baselines on diverse network settings. 
\end{abstract}


\section{Introduction}
Blockchain is rapidly emerging as a transformational technology for realizing trustless, decentralized and secure peer-to-peer (p2p) applications at large scales over the internet.
Cryptocurrency, the earliest proposed blockchain application, has grown to be a trillion dollar market today, while thousands of decentralized applications (dapps) are enjoying massive popularity in myriad domains including healthcare, social networks, decentralized web and beyond. 
The continued demand for cryptocurrencies and dapps, despite recent turbulence in market sentiments, also underscores the value of blockchains as protocols for realizing truly democratized applications~\cite{bitcoindemand}.

A blockchain is an append-only distributed ledger of transactions (e.g., payments) maintained by nodes of a p2p network. 
Nodes run a {\em consensus} protocol for ensuring consistent transaction ordering in the ledgers across the entire network. 
In Proof-of-Work (PoW) consensus---the prototypical blockchain consensus algorithm used by Bitcoin, Ethereum 1.0 and many other systems---transactions are packaged in to blocks that are regularly added to the blockchain through a process called mining.
A block is proposed (mined) by a node (miner) through solving a computationally difficult cryptographic puzzle, for which the miner receives a monetary fee as reward. 
To prevent Sybil attacks, the cryptographic puzzle difficulty is set so high today that miners use custom hardware within highly optimized mining farm facilities.   
The high cost of mining pits the miners in direct competition with each other, with each miner vying to mine as many blocks as possible and maximize profits. 

Miners resort to various strategies to increase their earnings, such as favoring locations with cheap electricity, or increasing hardware power consumption and cooling efficiencies of the mining farms~\cite{bitcoinelectricity,chow2017bitcoin,taylor2017evolution,stoll2019carbon}. 
They also organize themselves in to mining pools, apportioning mining rewards across pool miners for a steadier income over time. 
However, a fundamental factor affecting rewards earned by a miner is the latency of block propagation in the p2p network.
For a block mined by a miner to be included in the blockchain, the block must be propagated to other miners as quickly as possible through gossip over the p2p network. 
In the gossip process a mined block is immediately sent to the miner's neighbors, who then validate the block and forward the block to their neighbors and so on.  
Delayed block propagation causes the block to be `forked' which nulls the reward for the block. 
Similarly, blocks mined by other miners must also reach a miner fast to prevent mining on a forked chain. 

Propagating a block through a worldwide network of miners takes between 100s of milliseconds to a few seconds on popular blockchains today~\cite{croman2016scaling,decker2013information}.  
Speed of light delay over the vast distances between miners, and the number of hops blocks have to be relayed over in a sparse p2p network topology are major factors contributing to the propagation delay.  
A fresh block is mined once every few seconds on many blockchains---notably Ethereum, and Ethereum Classic which have a `block time' of 12 seconds on average~\cite{vujivcic2018blockchain}. 
The high delay in block propagation relative to block time means a speedup of even few 10s of milliseconds in a miner's network translates to significant gains in mining rewards for the miner over time~\cite{gao2019topology}.  

With a general understanding that faster the network the higher are the rewards earned, miners often subscribe to high-speed (few Gbps) Internet access links and make use of low-latency private backbone networks of cloud providers when possible~\cite{howtobuild,f2poolsec}.
Increasingly pool operators are also choosing private block relay networks (e.g., BloXroute~\cite{klarman2018bloxroute,bloxrouteweb}) for a more efficient block dispersion. 
However, recent work has shown that the dependence of mining rewards on propagation latency is more intricate than this~\cite{mao2022less}.    
Specifically, an honest miner that is well connected with other miners inadvertently creates efficient, low latency paths for other miners by acting as a centrally located bridge between the miners. 
However, to maximize the marginal gains in reward due to the network, it is important for a miner to have paths to other miners that are, on average, of a lower delay {\em relative} to the delays of paths between other miners. 
For example, if miners are arranged as a star topology with links of unit delay and uniform compute power across nodes, the central node receives a higher reward compared to the leaf nodes by including more blocks on the blockchain.  
On the other hand, on a complete graph topology with unit delay links and uniform compute power as before, all nodes receive the same reward.  
A node identically connected to other nodes in the two cases (i.e., the central node in the star topology and any arbitrary node in the complete graph topology: both have direct links to all other nodes) receives different rewards, as rewards depend not only on the node's own connections but also on how other nodes' connections. 
Thus, there is an inherent tension for a miner in increasing her own connectivity to the rest of the network while simultaneously ensuring that the connectivity between other miners do not significantly increase.  
A systematic research of this tension, and efficient connection policies to maximize marginal mining reward gain due to the network, have not been done to our best knowledge.   

In this work, we formalize the p2p topology construction problem as a game between miners and present \ourAlgorithm, a decentralized policy for optimizing reward. We consider a simplified setting where only a single node chooses its connections, while the rest of the network's topology is fixed. We assume that the global topology of the p2p network is unknown to miners. We thus model the problem of optimizing rewards by the connections-deciding miner node as a Markov decision process (MDP) with no state and an action set with a combinatorial number of actions. 

We derive the optimal neighbor selection policy using a combinatorial multi-armed bandit (MAB) approach \cite{chen2013combinatorial}. In the MAB algorithm, the agent (miner) explores various candidate connection configurations, and gradually adapts its connections based on past experience to gain the most mining rewards. 
A key contribution of our work is a network coordinates based model for efficiently learning the MAB environment~\cite{dabek2004vivaldi}. 
In this model, miners are assigned real-valued vectors from an Euclidean space, which capture the relative location of miners with respect to each other in the network.   
The coodinates are continuously updated based upon the reward feedback the agent receives from the environment. 
Thus, despite not having global knowledge of the network initially, we show that it is possible for an agent to learn about the network by just using the observed reward information. 

To enable the deployment of MAB algorithm, we have built a simulator. To simplify the reward computation in the simulator, rather than simulating the actual mining process at each step of the MDP, we consider a computationally easier function that only depends on the pairwise shortest path lengths between miners.  Importantly, our MDP reward function captures the property that a miner's mining gains depends on how small the shortest path lengths between the agent and other miners are relative to the shortest path lengths between other miners. 
Experimentally we show \ourAlgorithm~outperforms or matches heuristics on diverse network settings.

\section{Related Work}

P2P network design for optimizing mining rewards has remained a relatively under-explored topic in the community. 
The work that is closest to our is Perigee~\cite{mao2020perigee} which proposes an adaptive peer-selection algorithm for minimizing block propagation latency in the network. 
However, Perigee does not model the game-theoretic competition between miners. 
Subsequent  works~\cite{tang2022strategic,babel2022strategic} consider optimizing the network to maximize extractable value (MEV) from transactions.    
A number of prior works have exposed the impact of the network on mining~\cite{cao2021characterizing,gencer2018decentralization,kim2018measuring,park2019nodes,saad2019partitioning,wan2019evaluating,xiao2020modeling}.  
While these works generally suggest that better network connectivity translates to higher mining rewards earned, the competitive effects of network connectivity and methods to optimize them have not been discussed. 
Other related works include KadCast~\cite{rohrer2019kadcast} which proposes a Kadmila-based structured overlay for efficient block broadcast, and relay networks such as BloXroute~\cite{klarman2018bloxroute} for transports blocks quickly across vast geographic distances. 

The idea of network coordinates for p2p networks has been prominently explored in the network systems literature since the turn of the millenium, including distributed approaches to learn them~\cite{ledlie2006stable,ng2002predicting,dabek2004vivaldi}. 
More recently, a number of theoretical works have studied using low-distortion embeddings  in finite metrics (i.e., over finite graphs) for various applications, e.g., sparse spanner construction~\cite{abraham2005metric,chan2015new,cohen2020light,filtser2022hop}.  

Game theory of blockchains, especially at the consensus layer, has received considerable attention. 
For example, Lewenberg et al.~\cite{lewenberg2015bitcoin} use game theory to study how mining rewards can be shared across members of a mining pool. 
On the other hand, prior works have considered various network games outside the context of blockchains ~\cite{galeotti2010network,roughgarden2010algorithmic}. 
Our work is the first (to our best knowledge) to consider network games in blockchains.

\section{Problem Formulation}
Let us consider a complete directed graph $\ALP G = (\ALP V, \ALP E)$, where $\ALP V$ is the set of nodes and $\ALP E$ is the set of directed edges. Each node in the graph represents a mining server. The hash rate of the mining server $v$ is denoted by $H_v$. 
We use $\VEC H$ to denote the hash rate vector $\VEC H:= (H_v)_{v\in \ALP V}$. 
A directed edge $(v_1, v_2) \in \mathcal{E}$ represents a (TCP) link between the nodes $v_1, v_2$.\footnote{We assume if $(v_1, v_2) \in \mathcal{E}$ then $(v_2, v_1) \in \mathcal{E}$ for all $v_1, v_2 \in \mathcal{V}$.}
The directed edge represents that  node $v_1$ can send messages (e.g., transactions, blocks etc.) to $v_2$ as and when required by the protocol. 
The time take for a message sent from $v_1$ to reach $v_2$ along the link $(v_1, v_2)$ is denoted by $l(v_1,v_2)\geq 0$.
For $(v_1, v_2) \notin \mathcal{V}$, we let $l(v_1, v_2)$ denote the latency of sending a message from $v_1$ to $v_2$ had there been a link between $v_1$ and $v_2$.  
We refer to $L:=[l(v_1,v_2)]_{v_1,v_2}\in\Re_+^{\ALP V\times\ALP V}$ as the latency matrix. 
In practice, the latency matrix is nearly a symmetric matrix; asymmetry in the latency between two nodes can arise occasionally if the forward and reverse IP paths between the nodes have significantly different lengths.
We assume that the hash rate of the nodes is publicly known by all the nodes.  
For instance, by inspecting the frequency with which blocks are mined by different miners on the blockchain, one can estimate the relative hash power of different nodes. 
Each node $v$ has knowledge of latency $l(v, u)$ between itself  and other nodes $u \in \mathcal{V}$, but does not have knowledge of the latency between other pairs of nodes. 
To get an estimate of the latency $l(v, u)$ to a node $u$, the node $v$ can simply issue a ping to $u$ and measure the round-trip-time.

Let $t\in\Na$ denote a time slot (or a round number). 
The duration of each time slot is a design parameter, and could be multiple minutes or hours long depending on the block production rate in the blockchain. 
Based on the information available to each node $v$, node $v$ must decide on a certain number of nodes to connect to for the running the PoW consensus protocol. 
We assume each node can only connect to a maximum of $\delta\in\Na$ nodes. 
To make a new connection, a node first sends a connection request to the recipient who either accepts or rejects the request.  
Each node can have at most $\gamma\in\Na$ incoming connections at any point in time. The numbers $\delta$ and $\gamma$ are public information and part of the consensus protocol specification.

If every node determines the connections strategically in every time slot to increase its expected reward, it leads to a multi-agent game problem. Since games are generally more difficult to solve, we first simplify the problem by assuming that the network topology is fixed and only one node is strategically picking the connections to the other nodes. This leads to a more tractable Markov decision process, which we describe next.  

\subsection{Markov Decision Problem Formulation}
\label{s:mdpform}

We now formulate the node connection problem as an MDP from the point of view of an individual node with the rest of the network being fixed. As we will see, this is a MDP with no state and has a combinatorial action set. 

Let $v_0\in\ALP V$ be the node that needs to determine the connections to other nodes. 
Let $\mathcal{F} \subset \mathcal{E}$ denote the links in $\mathcal{G}$ that have been made by nodes other than $v_0$, i.e., $\mathcal{F} = \{ (u, u') \in \mathcal{E} : u \neq v_0 \}$. 
The edges in $\mathcal{F}$ denote the fixed part of the network. 
The set of edges in $\ALP F$ will be augmented with the connections picked by $v_0$. 

The MDP is parametrized by the hash rate vector $\VEC H$, the roundtrip delay matrix $L$, and the parameters $\delta$ and $\gamma$.  We let the complete parameters be denoted by $\theta := (\VEC H, L, \delta, \gamma)$.

The action of the node $v_0$ at time $t$ is a set $a_{S,t} = \{ u_{t,1}, u_{2,t}, \ldots, u_{\delta, t} \}$ of nodes in $\mathcal{V} \setminus \{v_0 \}$ to whom $v_0$ connects to. 

Once the action is picked, the neighbors of the node $v_0$ are determined for the time $t$. The set of edges in the network at time $t$ is $\ALP F_t:=\ALP F\cup \{(v_0,v): v\in \VEC a_t\}\subset\ALP E$ over which the consensus protocol runs. 

Let $\ALP P_t(v,v')$ denote the set of paths from node $v$ to $v'$ for $v' \in \mathcal{V} \setminus \{v_0 \}$. Each path $P\in \ALP P_t(v,v')\subset\ALP F_t$ is an ordered collection $((v,u_1),(u_1,u_2),\ldots,(u_k,v'))$. Define $\bar l(v,v')$ as
\begin{align}
\bar l(v,v') = \min_{P\in \ALP P_t(v,v')}\sum_{(u,u')\in P} l(u,u'),
\end{align}
which measures the latency for messages starting from node $v$ to reach node $v'$.

We let $\ind{\ALP V}$ denote a vector of all 1s of size $|\ALP V|$. 
Let $\ALP U_{v}$ denote a collection of subsets of nodes with aggregate hash power greater than 90\% of the total hash power in the network as follows:
\begin{align}
\ALP U_v(\ALP F_t; \theta) = \left\{\ALP U\subset\ALP V\setminus \{v\}: \sum_{u\in\ALP U} H_u \geq 0.9 \ind{\ALP V}^T \VEC H\right\}.
\end{align}
 
Let $A_v(\ALP F_t; \theta)$ denote the 90 percentile network latency for node $v$. This is determined by 
\begin{align}
A_v(\ALP F_t; \theta) = \min_{U \in \mathcal{U}_v(\mathcal{F}_t; \theta)} \max_{u\in U}\bar l(v,u). \label{eq:avdefn}
\end{align}

Define $\bar A(\ALP F_t; \theta)$ as the average 90-th percentile latency of the network:
\begin{align}
\bar A(\ALP F_t; \theta) = \frac{1}{|\ALP V|}\sum_{v\in\ALP V}  A_v(\ALP F_t; \theta). 
\end{align}
For a node $v$, the amount of cryptocurrency earned during the mining process is fairly complicated function of the network topology. To keep the problem tractable, we pick a simple reward function for the node that attempts to minimize the 90 percentile network latency in comparison to the average 90 percentile network latency of all the nodes. Accordingly, the reward of the node $v_0$ is given by
\begin{align}
R_v(\VEC a_t) = -\beta\frac{A_v(\ALP F_t; \theta)}{\bar A(\ALP F_t; \theta)}, \label{eq:reward}
\end{align}
where $\beta$ is a positive constant that depends on the blockchain protocol and the length of the timeslot. 
Eq.~\eqref{eq:reward} captures the intuition that a miner receives greater than its fair share of rewards if the miner has network connectivity that is on average better than the network connectivity of other miners. 
In practice the reward during a time slot can be computed by measuring the cumulative mining fees earned over the duration of the time slot. 
The goal of the node $v_0$ is to pick its neighbors at every time step strategically so that the long term average expected reward is maximized, where the randomization (if any) comes from the miner's neighbor selection policy. Thus, we arrive at an MDP with no state and a combinatorial number of actions.

\begin{figure*}[!ht]
     \centering
     \begin{subfigure}[b]{0.3\textwidth}
         \centering
         \includegraphics[width=\textwidth]{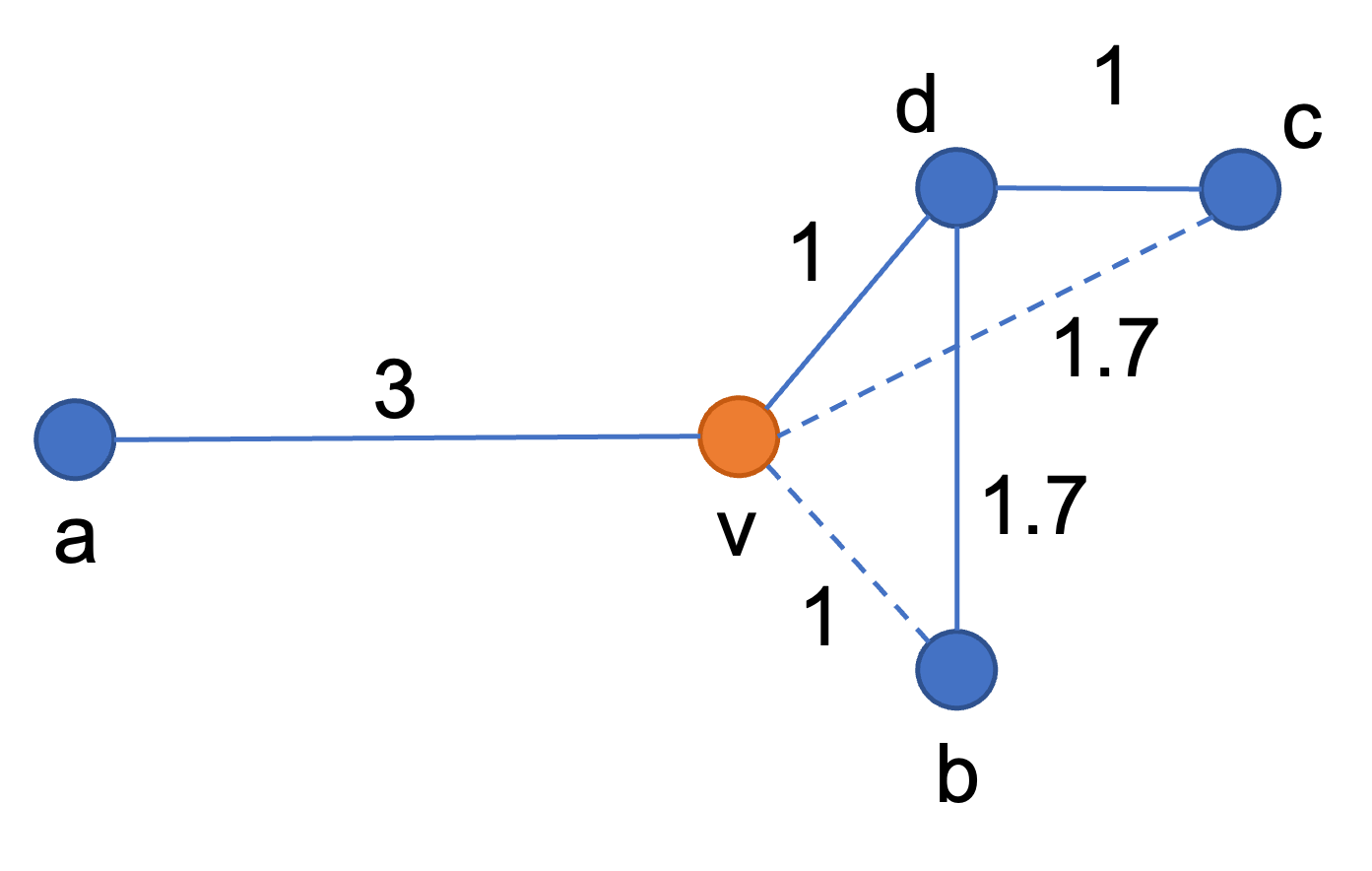}
         \caption{Base network}
         \label{fig:toyex1}
     \end{subfigure}
     \begin{subfigure}[b]{0.3\textwidth}
         \centering
         \includegraphics[width=\textwidth]{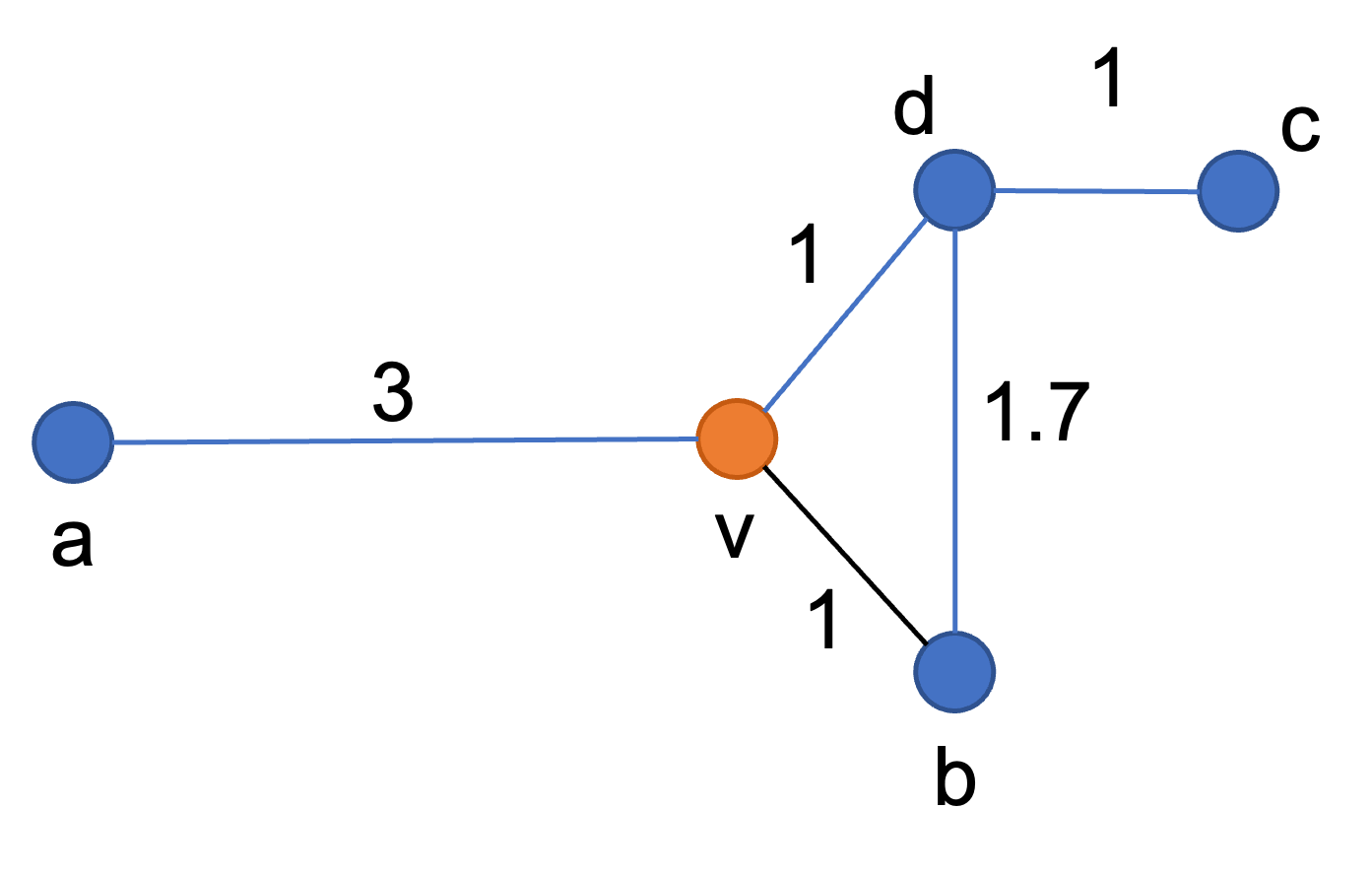}
         \caption{Player $v$ connects to node $b$}
         \label{fig:toyex2}
     \end{subfigure}
     \begin{subfigure}[b]{0.3\textwidth}
         \centering
         \includegraphics[width=\textwidth]{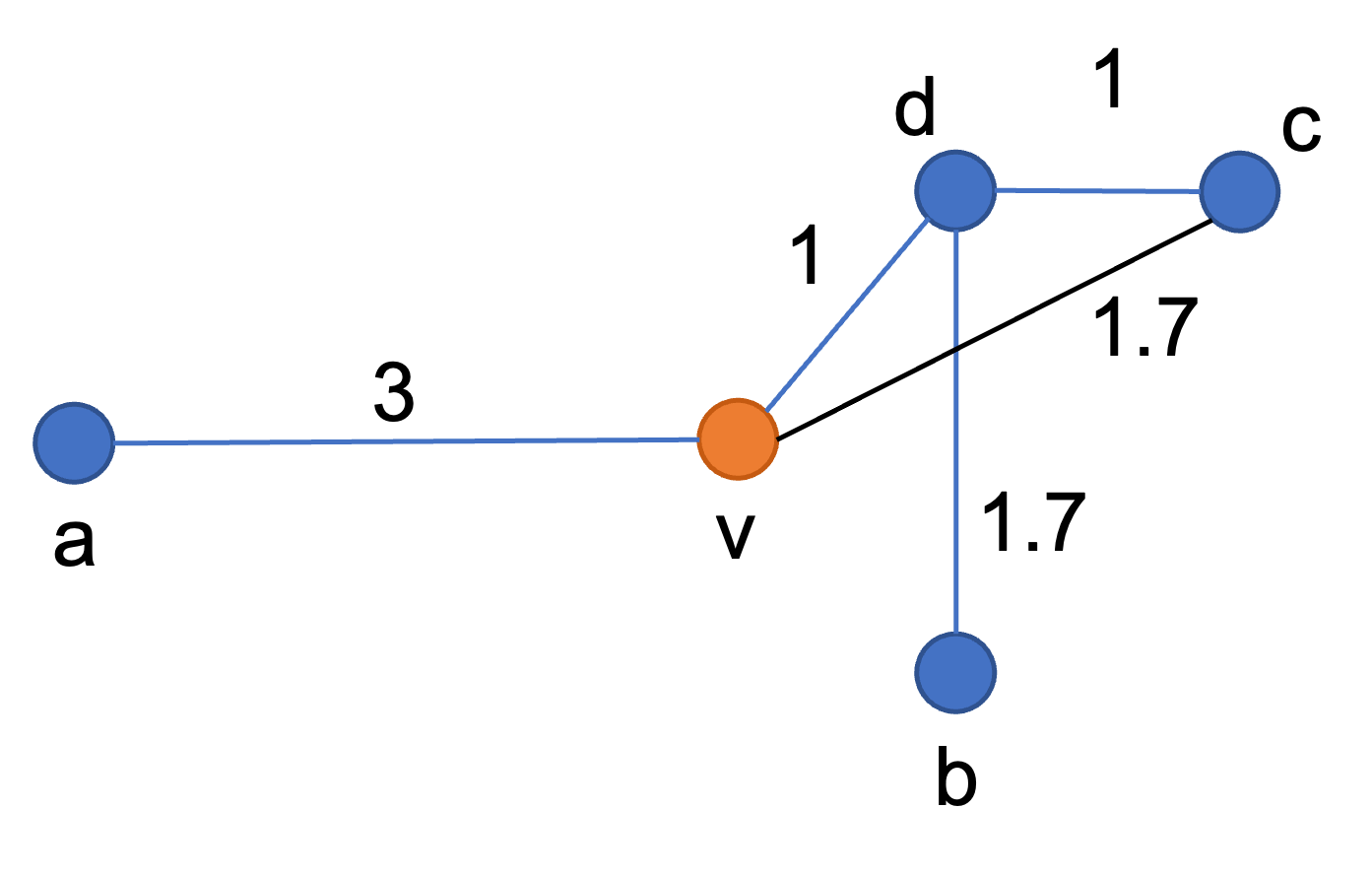}
         \caption{Player $v$ connects to node $c$}
         \label{fig:toyex3}
     \end{subfigure}     
        \caption{Example of a network showing the effect of player $v$'s action on reward.}
        \label{fig:toyex}
\end{figure*}

\subsection{Key Issues}
From a node's perspective, the hash rate vector of the network and the parameters $\delta$ and $\gamma$ are public knowledge and therefore known to the node. However, each node $v$ can only observe the latency $l(v,\cdot)$ to all other nodes in the network. Thus, the node does not know the latency other nodes in the network are observing. Moreover, the edges associated with the other nodes in the network are also unknown to the node $v$. Thus, each node $v$ has partial knowledge of $L$ and $\ALP F$. 

Since we have a stateless MDP and each node does not know some graph and network parameters, we arrive at a ``multi-agent combinatorial bandit problem". If every node is strategic in the network and picks their neighbors strategically, then the multi-agent combinatorial bandit problem is very difficult to solve. Thus, for simplicity, we assume that all nodes except $v_0$ are non strategic, due to which the network edges $\ALP F$ is fixed (though node $v_0$ still has only partial knowledge of $\ALP F$). The goal of this paper is to solve the combinatorial bandit problem associated with node $v_0$'s optimization problem.

\subsection{Computation of $A_v(\ALP F_t; \theta)$}
Note that a na\"ive computation of $A_v(\ALP F_t; \theta)$ (Eq.~\eqref{eq:avdefn}) requires combinatorial number of operations. 
In this section, we identify a simple algorithm which computes the value $A_v(\ALP F_t; \theta)$ in polynomial time.

\begin{algorithm}[h!]
\caption{Computing $A_v(\mathcal{F}_t;\theta)$ }
\label{alg:avcompute}
\begin{algorithmic}[1]
\STATE {\bf Input:} $v$, graph $\mathcal{F}_t$, latency matrix $L$, hash power $\mathbf{H}$
\STATE {\bf Output:} $A_v(\mathcal{F}_t; \theta)$
\STATE Compute $\bar l(v,v')$ for all $v, v' \in \mathcal{V}$ in graph $\mathcal{F}_t$ 
\STATE Sort nodes in $\ALP V \setminus \{v\}$ as $(v_1, v_2, \ldots, v_{n-1})$ such that $\bar l (v, v_1) \leq \bar l (v, v_2) \leq \ldots \bar l (v, v_{n-1})$. Here, $n = |\ALP V|$.
\STATE Initialize $k \leftarrow 1, s_k \leftarrow \bar l(v,v_k)$.
\WHILE {$H_v +\sum_{i=1}^{k+1} H_{v_i} \leq 0.9\ind{\ALP V}^T \VEC H$}
\STATE $s_{k+1} \leftarrow \max\{s_k,\bar l(v,v_{k+1})\}$
\STATE $k \leftarrow k + 1$
\ENDWHILE
\RETURN $\max\{s_k,\bar l(v,v_{k+1})\}$
\end{algorithmic}
\end{algorithm}

\begin{lemma}
Algorithm~\ref{alg:avcompute} computes $A_v(\ALP F_t; \theta)$ as defined in Eq.~\eqref{eq:avdefn}, and does so in polynomial time.
\end{lemma}

\section{Motivation}
Before we present our algorithm, we motivate the problem with a simple, toy example.
Consider a network of 5 nodes as shown in Fig.~\ref{fig:toyex1}, with node $v$ being the player node.  
The solid edges show existing links in the network, while dotted edges are potential links that can be added by the player if it chooses to. For each link, Fig.~\ref{fig:toyex1} also indicates the latency of the link. 
For simplicity, suppose player $v$ is interested in choosing $\delta=1$ new connection in the network. 
Further, in this example we define $A_v$ as the 100-th percentile latency for any node $v$, instead of the 90-th percentile latency as in Eq.~\eqref{eq:avdefn}.
The precise hash power distribution across nodes therefore does not matter for this example. 
The reward earned by the player is as in Eq.\eqref{eq:reward} with $\beta = 1$. 

Fig.~\ref{fig:toyex2} and~\ref{fig:toyex3} show two possible actions for player $v$. 
In Fig.~\ref{fig:toyex2} node $v$ makes the connection to node $b$, while in Fig.~\ref{fig:toyex3} node $v$ makes the connection to node $c$. 
If $v$ connects to $b$, the 100-th percentile latency values of nodes $a, b, c, d, v$ are $5, 4, 5, 4, 3$ respectively, which gives a reward of $-0.71$ to $v$. 
On the other hand, if $v$ connects to $c$, the 100-th percentile latency values of nodes $a, b, c, d, v$ are $5.7, 5.7, 4.7, 4, 3$ respectively, which now provides a reward of $-0.65$ to the player. 

Thus, the amount of reward earned by the player depends on the choice of connections made by the player. 
The problem is non-trivial even if the player has to choose just a single new connection. 
Intuitively, in Fig.~\ref{fig:toyex} no matter what action node $v$ chooses, its 100-th percentile latency remains the same. 
However, the choice of connections crucially impacts the 100-th percentile latency values of the other nodes in the network. 
For example, in Fig.~\ref{fig:toyex2}, while $v$'s connection to $b$ does not provide any benefit to $v$ it significantly shortens the path for $b$ to reach $a$, which ultimately reduces the reward earned by $v$. 
In Fig.~\ref{fig:toyex3}, however, by connecting to $c$, node $v$ ensures $b$'s path length to $a$ remains large thus incurring a greater reward than in Fig.~\ref{fig:toyex2}. 
The key observation with this example is that if a node is already well-connected in the network, then any additional connections made by the node must not significantly help other nodes in reducing their latency. 
With larger sized networks, non-uniform hash power distribution, and $\delta > 1$, the problem becomes even more complex.   

\section{\ourAlgorithm~Design}
\label{s:design}

We now present \ourAlgorithm, a combinatorial bandit algorithm for choosing efficient actions by the player. 
\ourAlgorithm~is a fully decentralized algorithm that miners can use to determine effective peers to connect with to maximize their mining rewards.
The algorithm is inspired by a line of theoretical works in the combinatorial bandit literature~\cite{chen2013combinatorial,gai2012combinatorial,gupta2021multi,gai2010learning,kveton2015tight,slivkins2019introduction}. 
However, prior works typically assume a "simple" and known underlying model (or, oracle) for the bandit environment which allows for efficient algorithms and tractable analysis.  
In our case, it is a priori not clear whether the blockchain p2p networking environment seen from the perspective of an individual miner can be fit reasonably to any of the existing combinatorial bandit models in the literature.
A core contribution of \ourAlgorithm~is an efficient model based on the idea of network coordinates~\cite{dabek2004vivaldi,ledlie2007network,cox2004practical,ng2002predicting,ledlie2006stable}, which we claim can be used to learn an effective representation of the bandit environment from past observations. 
The network-coordinate model can then be used to estimate rewards obtained through playing different actions, using which the best action to take can be inferred. 

In a network-coordinate model, each node $v \in \mathcal{V}$ of the networks is assigned a real-valued vector of coordinates $x_v \in \mathbb{R}^k$ in an Euclidean space (e.g., $k=5$ in our evaluations).  
A node's coordinates represents its location in the network, relative to other nodes. 
For instance, two nodes with a low round-trip-time delay to each other should have coordinates that are close to each other in Euclidean distance, and vice-versa.
Network coordinates are real-valued which makes them amenable to be learned via gradient descent techniques within auto-differentiation packages. 
Thus, purely based on past bandit action-reward information observed by a player, our model creates an estimate of the network structure and computes effective actions the player can take.
While IP addresses of miners may be known publicly (e.g., through blockchain monitoring services~\cite{ethernodes}), an IP address does not reveal fine-grained information about a miner's relative position in the network. 
Moreover, IP addresses can easily be spoofed and do not reveal the topological structure of the network. 
We remark that alternative network models are possible: for instance, we can try to learn the miner-to-miner link latency directly. 
This approach requires $O(n^2$ parameters ($n = |\mathcal{V}|$), as opposed to the $O(n)$ parameters in our proposed network coordinates model. 
Learned link latency values also do not reveal topological information. 

\begin{algorithm}[!t]
\caption{Algorithm template for Oracle at node $v$}
\label{alg:overview}
\begin{algorithmic}[1]
\STATE Initialize: Oracle model and its parameters $\theta = [x_1, x_2, \ldots, x_n]$ where $x_i$ is estimate of network coordinate for node $i$; exploration parameter $\epsilon$; 
\FOR {each time step $t$}
\STATE Sample $R$ uniformly randomly between $[0, 1]$; 
\IF {$R < \epsilon$} 
\STATE Choose action $a_v(t)$ randomly and observe rewards $R_v(t)$; 
\ELSE 
\STATE  Choose $a_v(t) \leftarrow \textsc{BestAction}(\theta)$ as the best action and observe reward $R_v(t)$; 
\ENDIF  
\STATE Predicted reward $ \leftarrow  \textsc{EstimateReward}(\theta, a_v(t))$
\STATE Update Oracle model parameters: $\theta \leftarrow \textsc{UpdateModel}(\theta, a_v(t), R_v(t))$ 
\ENDFOR 
\end{algorithmic}
\end{algorithm}

Algorithm~\ref{alg:overview} presents an overview of \ourAlgorithm. 
At each time step, based on the current set of node network coordinates the algorithm computes the best action (function \textsc{BestAction}) the player can take. 
Upon playing the recommended action, the player receives a reward from the bandit environment. 
The player compares the reward obtained against its estimated reward for the action (function \textsc{EstimateReward}), and uses the discrepancy between the two to revise its network coordinate model (function \textsc{UpdateModel}). 
To avoid over-fitting, we follow an $\epsilon-$greedy schedule for exploring random actions occasionally. 
We discuss the subroutines below. 

\smallskip
\noindent
\textbf{\textsc{BestAction}():}
For a candidate action $a_v$ with network coordinates $\theta$, the \textsc{EstimateReward}$(\theta, a_v)$ function computes an estimate of the reward earned if the player plays action $a_v$. 
To compute what is the best action that can be played, the \textsc{BestAction}$(\theta)$ function simply exhaustively calls the \textsc{EstimateReward}$(\theta, a_v)$ routine for all possible candidate actions $a_v$, and selects the action having the highest estimated reward. 
An exhaustive search is possible for small networks; for larger networks we can consider a random subsample of candidate actions or other local search heuristics to reduce complexity. 

\smallskip 
\noindent 
\textbf{\textsc{EstimateReward}():}
To compute an estimate of the reward given a candidate action $a_v$, we first estimate the link latency between any two nodes $u, u' \in \mathcal{V}$ as $\hat{l}(u, u') = || x_u - x_{u'} ||_2$ where $||\cdot ||_2$ is the L2 norm.
Next, we compute an estimate of the network topology $\hat{\mathcal{E}}$. 
For the player $v$, we let links from $v$ to nodes in $a_v$ be the set of links incident to $v$ in $\hat{\mathcal{E}}$.  
For nodes $u, u' \in \mathcal{V}, u, u' \neq v$, we heuristically estimate whether $(u, u') \in \hat{\mathcal{E}}$ based on $\theta$.
A simple heuristic, for instance, is to randomly decide whether $(u, u') \in \hat{\mathcal{E}}$. 
Here whether $(u, u') \in \hat{\mathcal{E}}$ is estimated once at time step 0, and fixed afterwards. 
More complex heuristics that also take in to account the current values $\theta$ of the network coordinates to estimate $\hat{\mathcal{E}}$ are also possible. 
We have adopted the random estimation heuristic in our evaluations. 
We also discuss advantages of knowing additional information about the true topology in our experimental results. 
With both the network topology $\hat{\mathcal{E}}$ and link latencies estimated, we can compute the  90-th percentile latency $A_u$ for all $u\in \mathcal{V}$ in the graph $(\mathcal{V}, \hat{\mathcal{E}})$ and hence the reward estimate $\hat{R}_v(a_v)$. 

\begin{table*}[!t]
    \centering
     \begin{tabularx}{\textwidth}{|X|X|X|X|X|X|X|X|X|}
          \hline
        \textbf{Algorithm} & \multicolumn{2}{|X|}{\textbf{Amsterdam}}& \multicolumn{2}{|X|}{\textbf{Atlanta}}&\multicolumn{2}{|X|}{\textbf{Shanghai}}& \multicolumn{2}{|X|}{\textbf{Tokyo}}  \\
     \hline
        \textbf{} &\textbf{$A_v$}&\textbf{Reward}&\textbf{$A_v$}&\textbf{Reward}&\textbf{$A_v$}&\textbf{Reward}&\textbf{$A_v$}&\textbf{Reward}  \\
        \hline
         Random choice&205.097&-1.299&205.662&-1.106&208.901&-1.494&200.159&-1.229\\
         \hline
         Least Latency&204.104&-1.114&204.308&-0.9663&204.344&-1.079&199.852&-1.211\\
         \hline
         Most Hash Power&204.11&-1.089&204.457&-1.022&208.572&-1.332&204.913&-1.433\\
         \hline
         Cobalt(Ours)&204.105&-1.04&204.298&-0.9674&205.003&-1.094&203.246&-1.16\\
         \hline
    \end{tabularx}
    \caption{Average Reward obtained under real world hash power distribution by player nodes in Europe, North America and Asia respectively.}
    \label{table1}
\end{table*}

\smallskip 
\noindent 
\textbf{\textsc{UpdateModel}():}
The last step during a round is to update the network coordinate values $\theta$ by comparing the estimated reward against the actual reward obtained. 
We define a loss function $\lambda(t) = (R_v(t) - \textsc{EstimateReward}(\theta, a_v(t)))^2$, where $R_v(t), a_v(t)$ denote the reward obtained and action taken during time step $t$, respectively. 
The coordinates are then updated as 
\begin{align}
\theta(t+1) \leftarrow \theta(t) - \eta \nabla_\theta \lambda(t),  
\end{align}
where $\eta > 0$ is a step-size parameter. 
Note that computing $\nabla_\theta \lambda(t)$ requires taking the gradient of \textsc{EstimateReward}$(\theta, a_v(t))$ with respect to $\theta$, which in turn requires computing the gradient of the 90-th percentile latency $A_u$ for $u \in \mathcal{V}$.  
This can easily be done as $A_u$ for any $u$ can be expressed as a sum of link latencies (estimated via network coordinates) along the 90-th percentlie latency path which is differentiable with respect to $\theta$. 

\section{Simulation Setup and Evaluation}
In this section, we illustrate the effectiveness of \ourAlgorithm~through experimental evaluation. 
We build a custom simulator on Python following the model described in \S\ref{s:mdpform}.

\subsection{Dataset Used}
A custom dataset has been generated using publicly available Ethereum blockchain data. Training dataset was created by selecting the major Ethereum Mining pools responsible for mining the most blocks, with their available network hash rates, and their estimated server locations~\cite{etherscan,ethmineloc}. The round trip times have been collected from publicly available ping data for major cities based on where these Ethereum nodes were estimated to be located~\cite{wondnetping}. Further synthetic distributions of hash powers have also been added to the training data, (we consider uniform hash rates as well as those drawn from an exponential distribution, apart from the true values). The ground-truth topology of the network is assumed to be static, and has been abstracted from the previous blocks of the Ethereum blockchain network\cite{etherscan}.

\subsection{Baselines}
We consider the following baselines in our evaluations:
\begin{itemize}
    \item Random Selection: In this, a playing node selects four connections randomly. This is the standard practice for a current node while connecting to a blockchain network.
    \item Least Latency: This heuristic involves a player node simply choosing the 4 neighbors that provide the lowest link latencies. 
    \item Most Hash Power: Similar to LeastLatency, the player node selects 4 nodes that provide the maximum hash rate.
\end{itemize}
We compare the above baselines to our slightly more strategic approach that takes a look at the 90 percentile network latency using a multi-armed bandit approach. 

\subsection{Implementation Details}
We run the experiment for $T=300$ rounds. We restrict the number of allowable connections to 4, that are chosen according to our neighbor selection policy. Here, $\beta=1$ and exploration parameter $\epsilon$=0.1. 
For a single node $v_0$, our action can only be to either connect to a new node after disconnecting from an existing node, or to keep the connection with an existing node, such that the total number of connections does not exceed 4 at any point of time. We initialize our network coordinates for the nodes randomly, and assume a random topology for the Oracle as described in \S\ref{s:design}. 
The coordinate system used here is a five dimensional one. 
We use the coordinates for estimating reward as described in \S\ref{s:design}.

\subsection{Results}
Table~\ref{table1} displays our results for the listed method compared to baseline heuristics. We observe the results for different initializations of $v_0$ being the cities Amsterdam, Atlanta, Shanghai and Tokyo which are some of the major hubs for mining Ethereum blocks in the North American, European and Asian continents respectively. 

\begin{figure}
     \centering
     \begin{subfigure}[b]{0.5\textwidth}
         \centering
         \includegraphics[width=\textwidth]{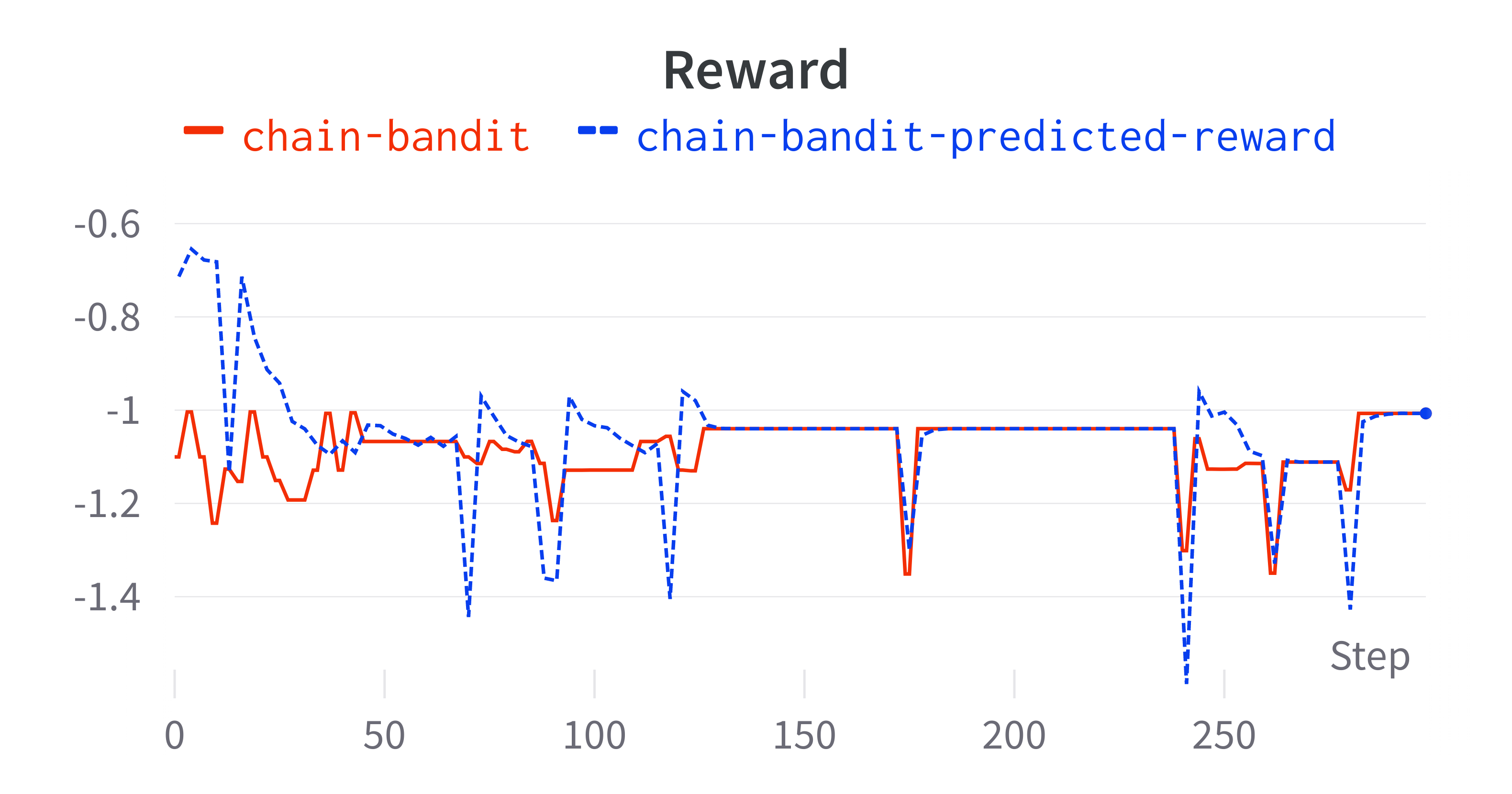}
         \caption{Player node Amsterdam}
         \label{fig4}
     \end{subfigure}\\
     \begin{subfigure}[b]{0.5\textwidth}
         \centering
         \includegraphics[width=\textwidth]{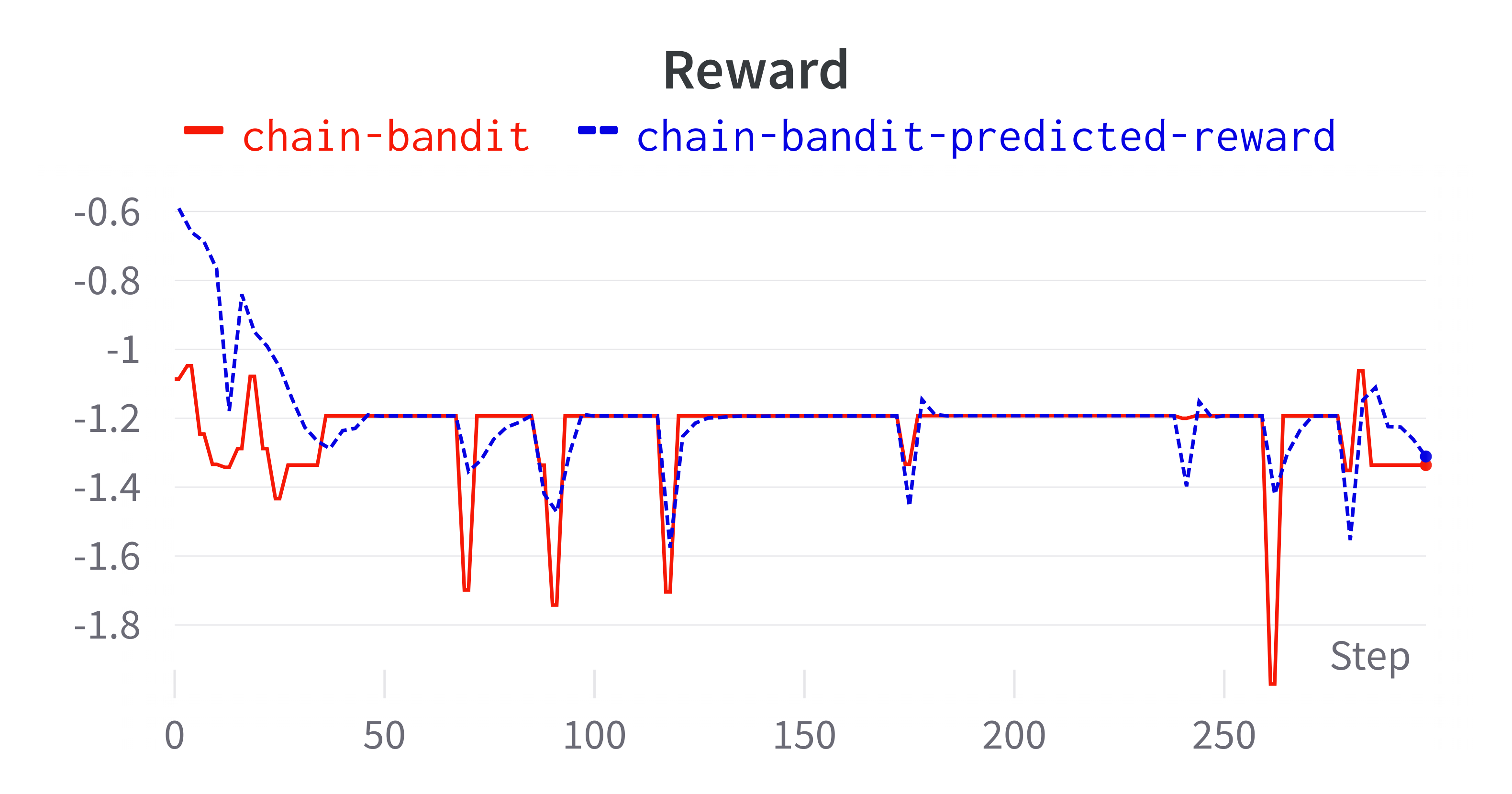}
         \caption{Player Node Shanghai}
         \label{fig5}
     \end{subfigure}
        \caption{Prediction estimates of reward obtained by well-connected and poorly connected players and their actual rewards obtained.}
        \label{fig:predictvsactual}
\end{figure}

We perform detailed experiments on player nodes Shanghai and Amsterdam, where Amsterdam is considered a "well connected" node and Shanghai, a "poorly connected" node. 

\subsubsection{Real world hash}
In current blockchain networks, we observe that most of the hash power is concentrated in a few key regions (greater than 50\% in the United States and Europe for Ethereum nodes \cite{ethernodes}) , whereas the remainder is distributed among other scattered nodes. As such, a node located in a region without this concentration of hash power (such as Asia), may face many challenges to improving their mining rewards as compared to a node located in a well connected region (such as North America). One of the solutions of this problem in a topology like this for a node wishing to increase their rewards without heavily investing into additional infrastructure would be to shift away from randomly selecting their neighbors and picking their connections more strategically, by following some kind of policy to choose the nodes. \ref{table1} affirms that this is indeed a natural next step, as the rewards obtained by random selection are always suboptimal. 

\begin{figure}
     \centering
     \begin{subfigure}[b]{0.5\textwidth}
         \centering
         \includegraphics[width=\textwidth]{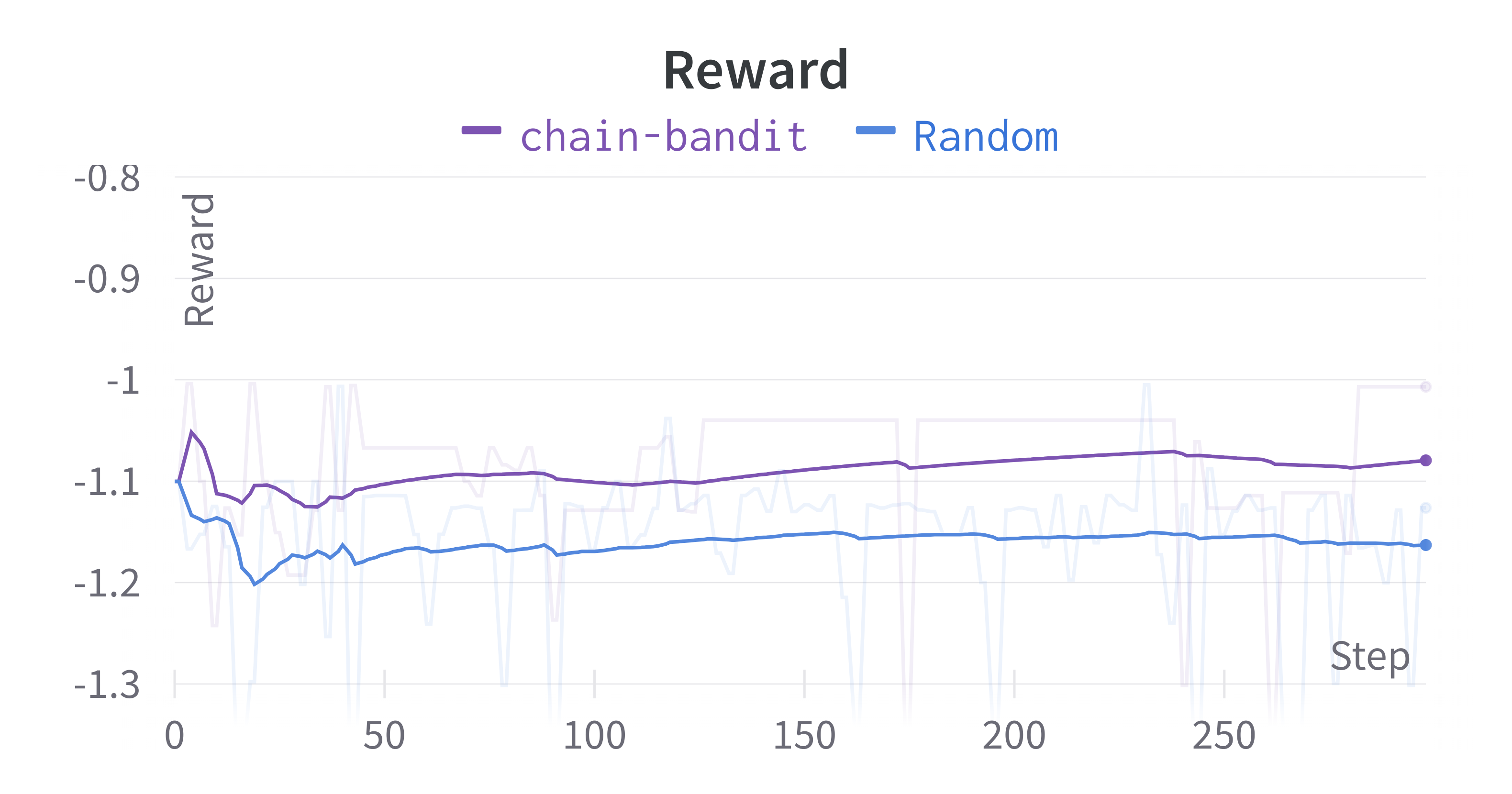}
         \caption{Player node Amsterdam}
         \label{fig1}
     \end{subfigure}\\
     \begin{subfigure}[b]{0.5\textwidth}
         \centering
         \includegraphics[width=\textwidth]{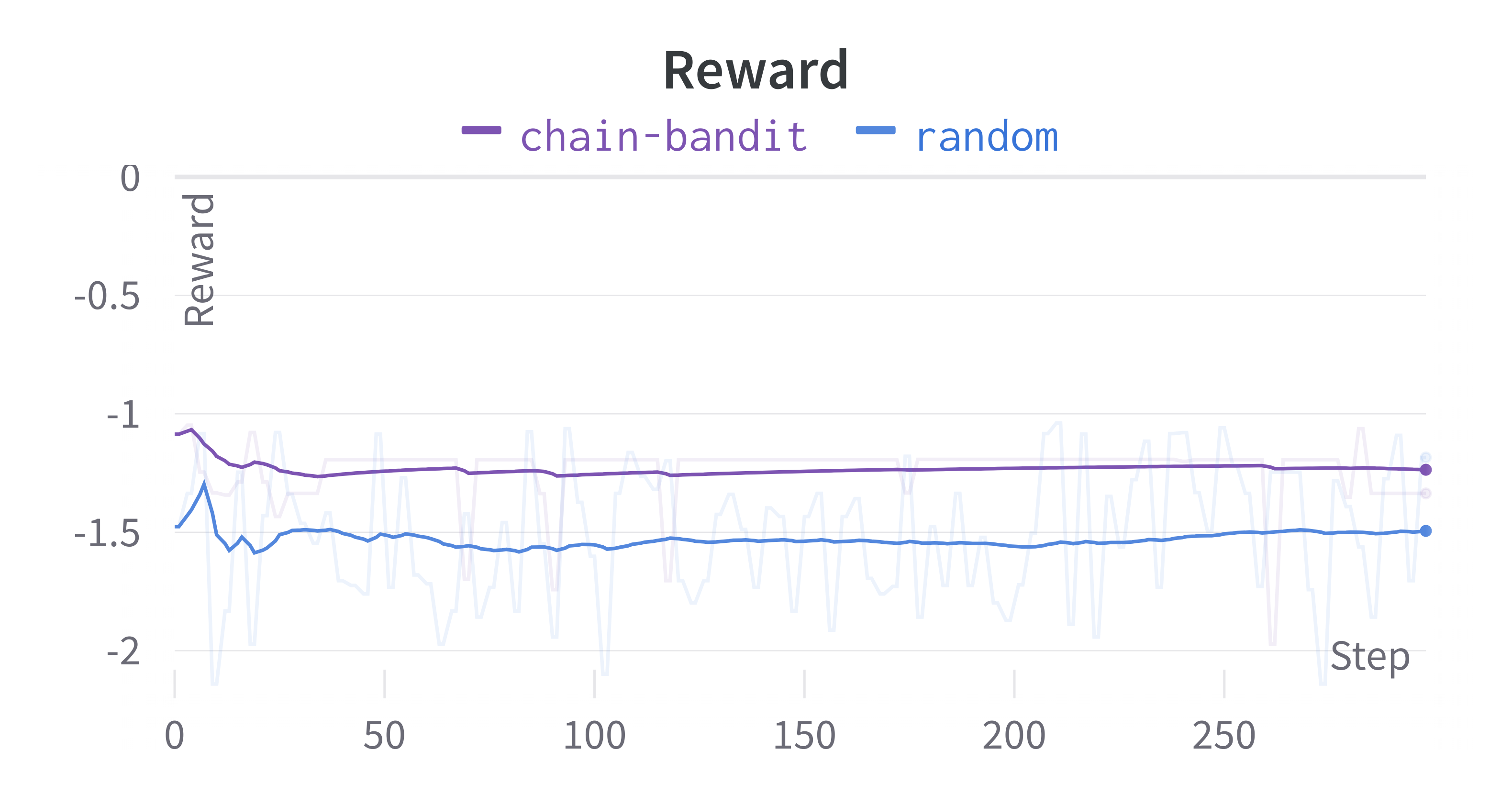}
         \caption{Player Node Shanghai}
         \label{fig2}
     \end{subfigure}
        \caption{Reward obtained by well-connected and poorly connected players v/s random selection under uniform hash power.}
        \label{playervsrandom}
\end{figure}

In this topology motivated by the real world, Fig.~\ref{fig:predictvsactual} describes the actual rewards obtained by a player as well as the prediction curve by our algorithm on both well connected and poorly connected players, and we see that for a complex network, our model is able to successfully learn the optimal action to take to converge to the best rewards. 

\subsubsection{Uniform hash and exponential hash}
Apart from the real world scenario, we also look at other variations of topologies, starting with the ideal case, where we assume that all nodes have equal hash power. Here too (Fig.~\ref{playervsrandom}), it is reasonable to expect to outperform random selection for both well connected and poorly connected nodes, simply by virtue of the playing node
being unable to select the optimal arm each time. However, here as our problem has been simplified, the only parameter that matters to the playing node becomes the link latencies, and the optimal action would be to connect to the neighbors closest to it. Indeed, our results reflect this case, and the best action selected by our algorithm eventually converges to the least latency arm.

\begin{figure}
     \centering
     \begin{subfigure}[b]{0.5\textwidth}
         \centering
         \includegraphics[width=\textwidth]{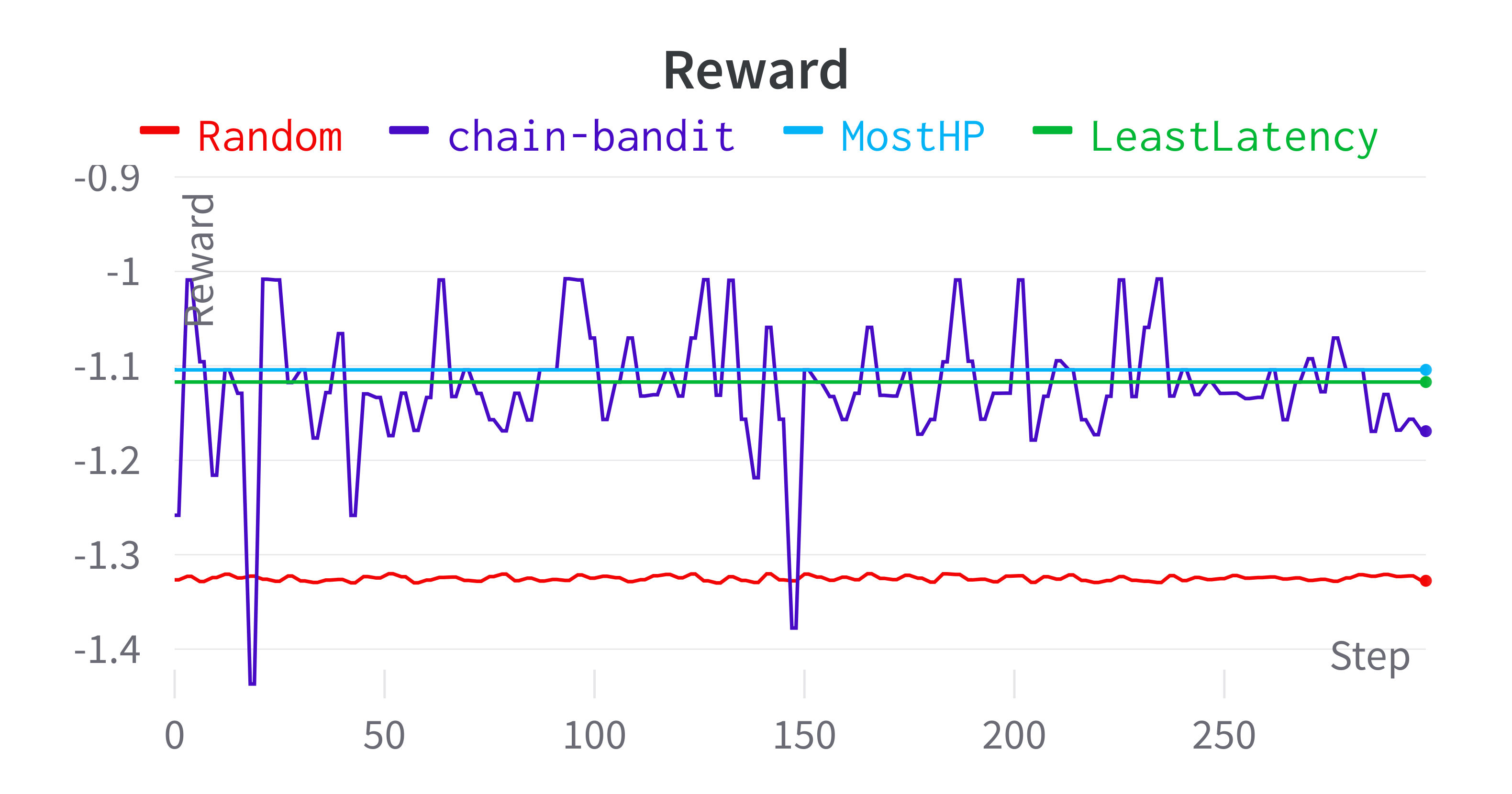}
         \caption{Player node Amsterdam}
         \label{fig7}
     \end{subfigure}\\
     \begin{subfigure}[b]{0.5\textwidth}
         \centering
         \includegraphics[width=\textwidth]{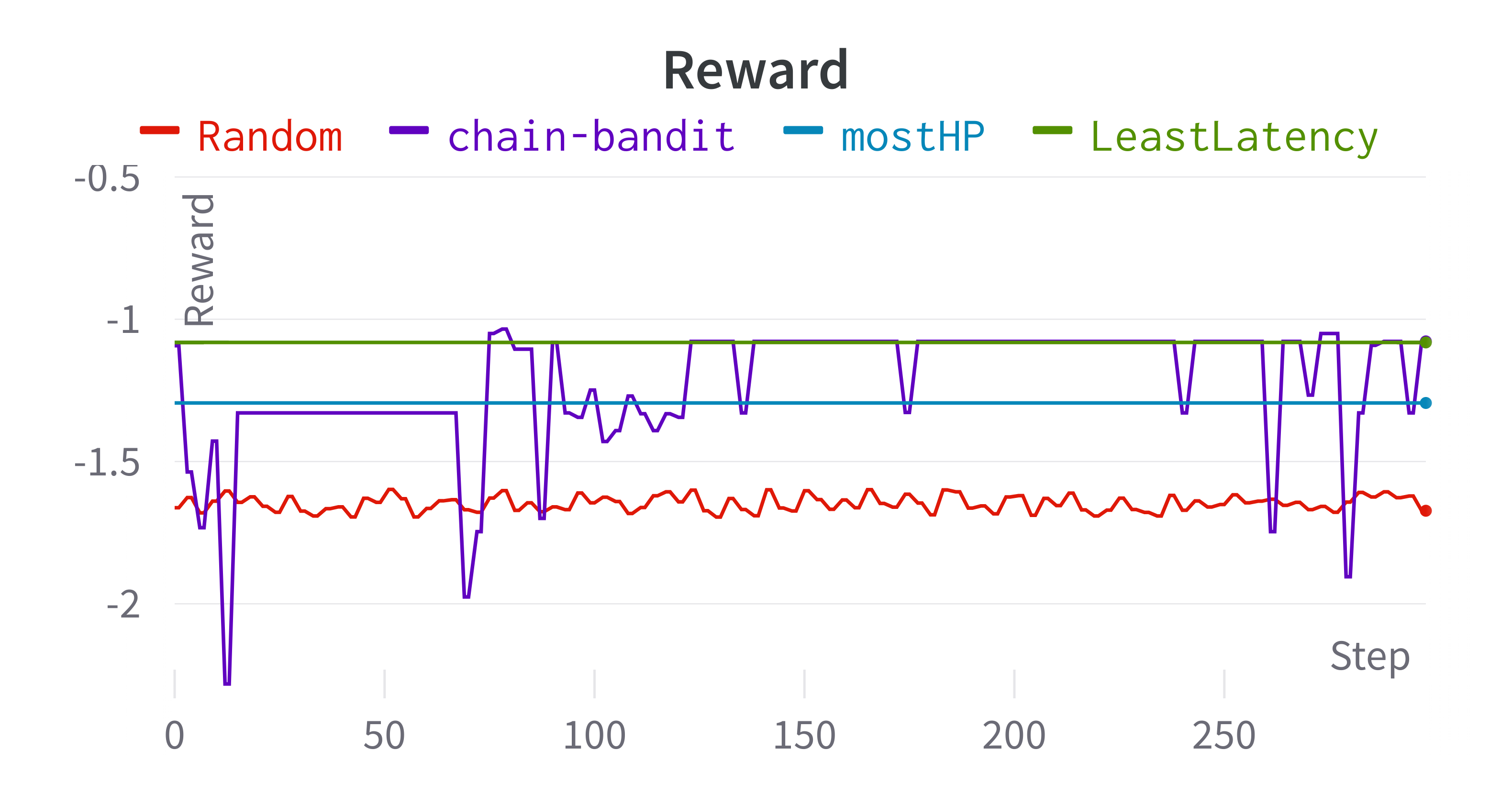}
         \caption{Player Node Shanghai}
         \label{fig8}
     \end{subfigure}
        \caption{Reward obtained by well-connected and poorly connected players for all heuristics vs chain bandits for an exponentially distributed hash power setting}
        \label{rewarddistexp}
\end{figure}

A slightly more complex example would be if we vary the hash rate. Here, we use data that contains nodes with hash powers that have been drawn from an exponential distribution. Fig. ~\ref{rewarddistexp} suggests that for a playing node, simple heuristics might be a better choice from the perspective of ease of use, however \ourAlgorithm~is very competitive in terms of the rewards obtained. Picking the least latency heuristic might be the cheapest in case of a network where there is a certain degree of confidence that the hash rates are unequal, but not very disparate from each other. For such a kind of network topology, a prudent approach could be to consider a simpler heuristic.  

\subsubsection{Role of topology estimate in Oracle}
To compute a reliable estimate of the reward function given a candidate action, Algorithm~\ref{alg:overview} requires an estimate of the underlying network topology.
In the results so far, we have assumed a random estimate of the topology. 

\begin{figure}[]
     \centering
     \begin{subfigure}[b]{0.45\textwidth}
         \centering
         \includegraphics[width=\textwidth]{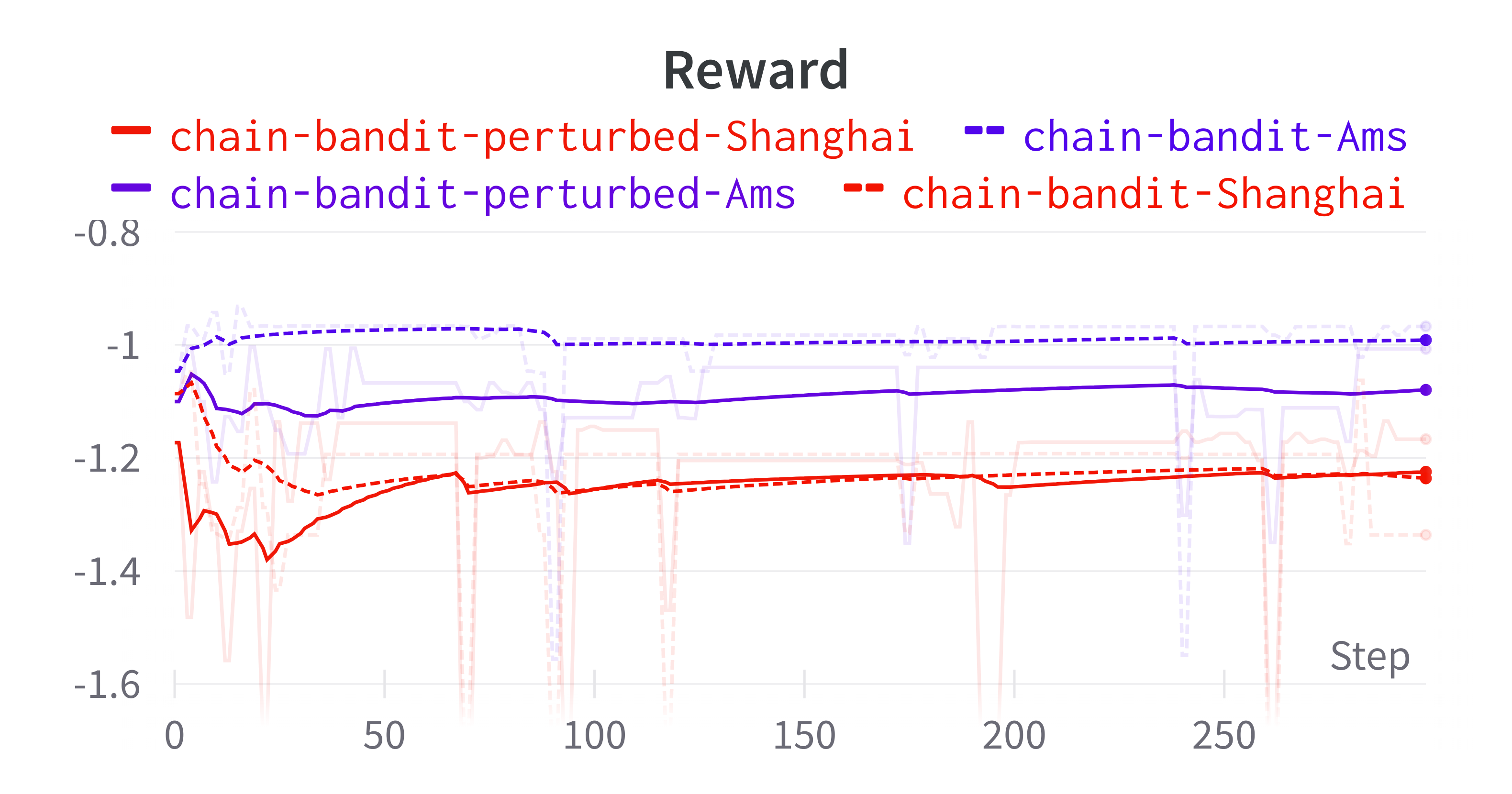}
         \caption{}
         \label{fig10}
     \end{subfigure}\\
     \begin{subfigure}[b]{0.45\textwidth}
         \centering
         \includegraphics[width=\textwidth]{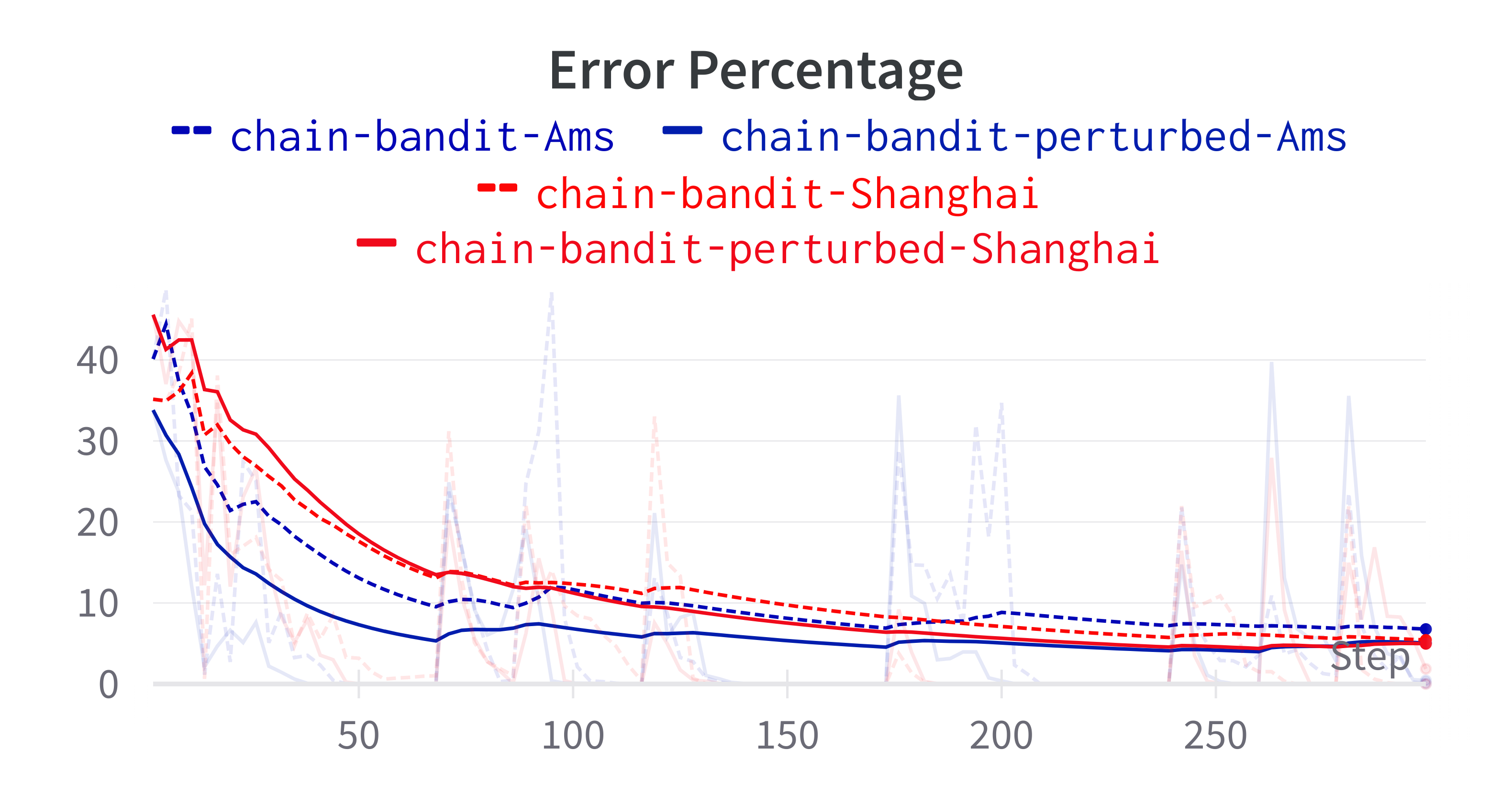}
         \caption{}
         \label{fig11}
     \end{subfigure}
        \caption{Reward and error curves under random and exact topology estimates in the Oracle. Random topology estimate is referred to as `perturbed' in the legend.} 
        \label{gndvsperturbed}
\end{figure}

From the perspective of a playing node, knowing the quality of active links between its neighbors is very useful in order to estimate the delays between the blocks being received by this node in the blockchain network, but typically, the underlying topology of the network is unknown to a player node, as a result learning a good action makes more sense. Fig.~\ref{gndvsperturbed} shows that our algorithm learns improved policies if the topology estimate used in the Oracle matches with the actual topology of the network. 

\begin{figure}[]
     \centering
     \begin{subfigure}[b]{0.45\textwidth}
         \centering
         \includegraphics[width=\textwidth]{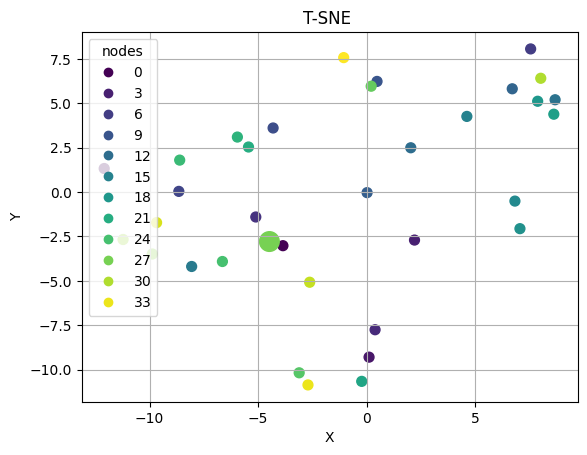}
         \caption{Initial Network Coordinates}
         \label{fig1}
     \end{subfigure}\\
     \begin{subfigure}[b]{0.45\textwidth}
         \centering
         \includegraphics[width=\textwidth]{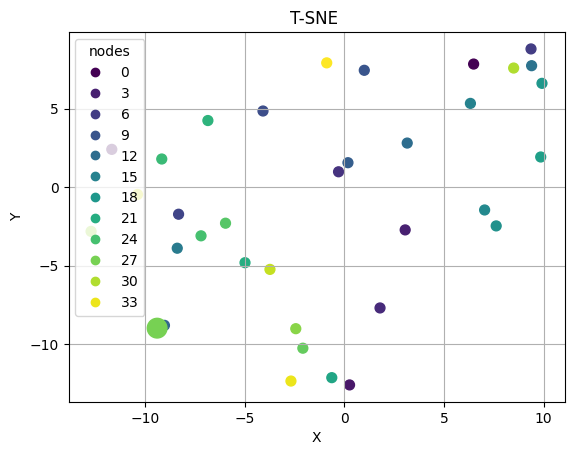}
         \caption{Final Network Coordinates}
         \label{fig2}
     \end{subfigure}
        \caption{t-SNE visualization of initial and final Network Coordinates for player Shanghai}
        \label{tSNE}
\end{figure}

\subsubsection{Interpreting the network coordinates}
Fig.~\ref{tSNE} shows the t-SNE visualization\cite{van2008visualizing} of the network coordinates that have been estimated by our model. We start with a random model, however as time passes, we observe our model trying to mimic the topology of the ground truth coordinates. 
For instance, we observe that even though the coordinates for Shanghai (shown by the large green point in Fig.~\ref{tSNE}) starts out being the middle, as time progresses the Shanghai's coordinates tends to move out to the periphery which is consistent with Shanghai's remote location relative to other nodes in our dataset.

\section{Conclusion}
In this paper, we have introduced \ourAlgorithm, a network coordinates based model that maximizes mining rewards earned by a miner through careful choice of neighbor connections on the P2P topology. 
We modeled the problem as a combinatorial bandit problem in which a node has only partial knowledge of the rest of the network 
and show that it is useful for a playing node to consider a careful selection of neighboring nodes instead of randomly choosing their connections. 

\subsection{Future Work}
Our future work involves expanding this to a dynamic system, where all nodes are allowed to choose their neighbors and we attempt to understand the equilibrium of this game problem.

Another future direction is the development of other network topology estimates for the oracle. Our current model uses a random estimate, however others (such as a coordinate dependent topology estimate) and their impact on the model would be an interesting study.

\bibliographystyle{abbrv}
\bibliography{references}

\begin{thebibliography}{10}

\bibitem{bloxrouteweb}
Bloxroute.
\newblock \url{https://bloxroute.com/}.

\bibitem{ethmineloc}
Ethereum mining locations.
\newblock \url{https://investoon.com/charts/mining-map/eth}.

\bibitem{ethernodes}
Ethernodes.
\newblock \url{https://ethernodes.org/}.

\bibitem{etherscan}
Etherscan.
\newblock \url{https://etherscan.io/}.

\bibitem{howtobuild}
How to build an {E}thereum mining pool.
\newblock
  \url{https://medium.com/dragonfly-research/how-to-build-an-ethereum-mining-pool-6be356520b7a}.

\bibitem{f2poolsec}
The secret weapon f2pool used to tackle its uncle rate.
\newblock
  \url{https://medium.com/bloxroute/the-secret-weapon-f2pool-used-to-tackle-its-uncle-rate-1ecb6fe47ef8}.

\bibitem{wondnetping}
Wondernetwork ping statistics.
\newblock \url{https://wondernetwork.com/}.

\bibitem{bitcoindemand}
The crypto market crashed. they’re still buying bitcoin., 2022.
\newblock
  \url{https://www.nytimes.com/2022/08/02/technology/crypto-bitcoin-maximalists.html}.

\bibitem{bitcoinelectricity}
This map shows the best states for bitcoin mining, 2022.
\newblock
  \url{https://www.cnbc.com/2021/09/30/this-map-shows-the-best-us-states-to-mine-for-bitcoin.html}.

\bibitem{abraham2005metric}
I.~Abraham, Y.~Bartal, J.~Kleinberg, T.-H. Chan, O.~Neiman, K.~Dhamdhere,
  A.~Slivkins, and A.~Gupta.
\newblock Metric embeddings with relaxed guarantees.
\newblock In {\em 46th Annual IEEE Symposium on Foundations of Computer Science
  (FOCS'05)}, pages 83--100. IEEE, 2005.

\bibitem{babel2022strategic}
K.~Babel and L.~Baker.
\newblock Strategic peer selection using transaction value and latency.
\newblock In {\em Proceedings of the 2022 ACM CCS Workshop on Decentralized
  Finance and Security}, pages 9--14, 2022.

\bibitem{cao2021characterizing}
T.~Cao, J.~Decouchant, J.~Yu, and P.~Esteves-Verissimo.
\newblock Characterizing the impact of network delay on bitcoin mining.
\newblock In {\em 2021 40th International Symposium on Reliable Distributed
  Systems (SRDS)}, pages 109--119. IEEE, 2021.

\bibitem{chan2015new}
T.-H.~H. Chan, M.~Li, L.~Ning, and S.~Solomon.
\newblock New doubling spanners: Better and simpler.
\newblock {\em SIAM Journal on Computing}, 44(1):37--53, 2015.

\bibitem{chen2013combinatorial}
W.~Chen, Y.~Wang, and Y.~Yuan.
\newblock Combinatorial multi-armed bandit: General framework and applications.
\newblock In {\em International conference on machine learning}, pages
  151--159. PMLR, 2013.

\bibitem{chow2017bitcoin}
S.~Chow and M.~E. Peck.
\newblock The bitcoin mines of {C}hina.
\newblock {\em IEEE Spectrum}, 54(10):46--53, 2017.

\bibitem{cohen2020light}
V.~Cohen-Addad, A.~Filtser, P.~N. Klein, and H.~Le.
\newblock On light spanners, low-treewidth embeddings and efficient traversing
  in minor-free graphs.
\newblock In {\em 2020 IEEE 61st Annual Symposium on Foundations of Computer
  Science (FOCS)}, pages 589--600. IEEE, 2020.

\bibitem{cox2004practical}
R.~Cox, F.~Dabek, F.~Kaashoek, J.~Li, and R.~Morris.
\newblock Practical, distributed network coordinates.
\newblock {\em ACM SIGCOMM Computer Communication Review}, 34(1):113--118,
  2004.

\bibitem{croman2016scaling}
K.~Croman, C.~Decker, I.~Eyal, A.~E. Gencer, A.~Juels, A.~Kosba, A.~Miller,
  P.~Saxena, E.~Shi, E.~G. Sirer, et~al.
\newblock On scaling decentralized blockchains.
\newblock In {\em International conference on financial cryptography and data
  security}, pages 106--125. Springer, 2016.

\bibitem{dabek2004vivaldi}
F.~Dabek, R.~Cox, F.~Kaashoek, and R.~Morris.
\newblock Vivaldi: A decentralized network coordinate system.
\newblock {\em ACM SIGCOMM Computer Communication Review}, 34(4):15--26, 2004.

\bibitem{decker2013information}
C.~Decker and R.~Wattenhofer.
\newblock Information propagation in the bitcoin network.
\newblock In {\em IEEE P2P 2013 Proceedings}, pages 1--10. IEEE, 2013.

\bibitem{filtser2022hop}
A.~Filtser.
\newblock Hop-constrained metric embeddings and their applications.
\newblock In {\em 2021 IEEE 62nd Annual Symposium on Foundations of Computer
  Science (FOCS)}, pages 492--503. IEEE, 2022.

\bibitem{gai2010learning}
Y.~Gai, B.~Krishnamachari, and R.~Jain.
\newblock Learning multiuser channel allocations in cognitive radio networks: A
  combinatorial multi-armed bandit formulation.
\newblock In {\em 2010 IEEE Symposium on New Frontiers in Dynamic Spectrum
  (DySPAN)}, pages 1--9. IEEE, 2010.

\bibitem{gai2012combinatorial}
Y.~Gai, B.~Krishnamachari, and R.~Jain.
\newblock Combinatorial network optimization with unknown variables:
  Multi-armed bandits with linear rewards and individual observations.
\newblock {\em IEEE/ACM Transactions on Networking}, 20(5):1466--1478, 2012.

\bibitem{galeotti2010network}
A.~Galeotti, S.~Goyal, M.~O. Jackson, F.~Vega-Redondo, and L.~Yariv.
\newblock Network games.
\newblock {\em The review of economic studies}, 77(1):218--244, 2010.

\bibitem{gao2019topology}
Y.~Gao, J.~Shi, X.~Wang, Q.~Tan, C.~Zhao, and Z.~Yin.
\newblock Topology measurement and analysis on ethereum p2p network.
\newblock In {\em 2019 IEEE Symposium on Computers and Communications (ISCC)},
  pages 1--7. IEEE, 2019.

\bibitem{gencer2018decentralization}
A.~E. Gencer, S.~Basu, I.~Eyal, R.~v. Renesse, and E.~G. Sirer.
\newblock Decentralization in bitcoin and ethereum networks.
\newblock In {\em International Conference on Financial Cryptography and Data
  Security}, pages 439--457. Springer, 2018.

\bibitem{gupta2021multi}
S.~Gupta, S.~Chaudhari, G.~Joshi, and O.~Ya{\u{g}}an.
\newblock Multi-armed bandits with correlated arms.
\newblock {\em IEEE Transactions on Information Theory}, 67(10):6711--6732,
  2021.

\bibitem{kim2018measuring}
S.~K. Kim, Z.~Ma, S.~Murali, J.~Mason, A.~Miller, and M.~Bailey.
\newblock Measuring ethereum network peers.
\newblock In {\em Proceedings of the Internet Measurement Conference 2018},
  pages 91--104, 2018.

\bibitem{klarman2018bloxroute}
U.~Klarman, S.~Basu, A.~Kuzmanovic, and E.~G. Sirer.
\newblock bloxroute: A scalable trustless blockchain distribution network
  whitepaper.
\newblock {\em IEEE Internet of Things Journal}, 2018.

\bibitem{kveton2015tight}
B.~Kveton, Z.~Wen, A.~Ashkan, and C.~Szepesvari.
\newblock Tight regret bounds for stochastic combinatorial semi-bandits.
\newblock In {\em Artificial Intelligence and Statistics}, pages 535--543.
  PMLR, 2015.

\bibitem{ledlie2007network}
J.~Ledlie, P.~Gardner, and M.~I. Seltzer.
\newblock Network coordinates in the wild.
\newblock In {\em NSDI}, volume~7, pages 299--311, 2007.

\bibitem{ledlie2006stable}
J.~Ledlie, P.~Pietzuch, and M.~Seltzer.
\newblock Stable and accurate network coordinates.
\newblock In {\em 26th IEEE International Conference on Distributed Computing
  Systems (ICDCS'06)}, pages 74--74. IEEE, 2006.

\bibitem{lewenberg2015bitcoin}
Y.~Lewenberg, Y.~Bachrach, Y.~Sompolinsky, A.~Zohar, and J.~S. Rosenschein.
\newblock Bitcoin mining pools: A cooperative game theoretic analysis.
\newblock In {\em Proceedings of the 2015 international conference on
  autonomous agents and multiagent systems}, pages 919--927. Citeseer, 2015.

\bibitem{mao2020perigee}
Y.~Mao, S.~Deb, S.~B. Venkatakrishnan, S.~Kannan, and K.~Srinivasan.
\newblock Perigee: Efficient peer-to-peer network design for blockchains.
\newblock In {\em Proceedings of the 39th Symposium on Principles of
  Distributed Computing}, pages 428--437, 2020.

\bibitem{mao2022less}
Y.~Mao and S.~B. Venkatakrishnan.
\newblock Less is more: Fairness in wide-area proof-of-work blockchain
  networks.
\newblock {\em arXiv preprint arXiv:2204.02461}, 2022.

\bibitem{ng2002predicting}
T.~E. Ng and H.~Zhang.
\newblock Predicting internet network distance with coordinates-based
  approaches.
\newblock In {\em Proceedings. Twenty-First Annual Joint Conference of the IEEE
  Computer and Communications Societies}, volume~1, pages 170--179. IEEE, 2002.

\bibitem{park2019nodes}
S.~Park, S.~Im, Y.~Seol, and J.~Paek.
\newblock Nodes in the bitcoin network: Comparative measurement study and
  survey.
\newblock {\em IEEE Access}, 7:57009--57022, 2019.

\bibitem{rohrer2019kadcast}
E.~Rohrer and F.~Tschorsch.
\newblock Kadcast: A structured approach to broadcast in blockchain networks.
\newblock In {\em Proceedings of the 1st ACM Conference on Advances in
  Financial Technologies}, pages 199--213, 2019.

\bibitem{roughgarden2010algorithmic}
T.~Roughgarden.
\newblock Algorithmic game theory.
\newblock {\em Communications of the ACM}, 53(7):78--86, 2010.

\bibitem{saad2019partitioning}
M.~Saad, V.~Cook, L.~Nguyen, M.~T. Thai, and A.~Mohaisen.
\newblock Partitioning attacks on bitcoin: Colliding space, time, and logic.
\newblock In {\em 2019 IEEE 39th international conference on distributed
  computing systems (ICDCS)}, pages 1175--1187. IEEE, 2019.

\bibitem{slivkins2019introduction}
A.~Slivkins et~al.
\newblock Introduction to multi-armed bandits.
\newblock {\em Foundations and Trends{\textregistered} in Machine Learning},
  12(1-2):1--286, 2019.

\bibitem{stoll2019carbon}
C.~Stoll, L.~Klaa{\ss}en, and U.~Gallersd{\"o}rfer.
\newblock The carbon footprint of bitcoin.
\newblock {\em Joule}, 3(7):1647--1661, 2019.

\bibitem{tang2022strategic}
W.~Tang, L.~Kiffer, G.~Fanti, and A.~Juels.
\newblock Strategic latency reduction in blockchain peer-to-peer networks.
\newblock {\em arXiv preprint arXiv:2205.06837}, 2022.

\bibitem{taylor2017evolution}
M.~B. Taylor.
\newblock The evolution of bitcoin hardware.
\newblock {\em Computer}, 50(9):58--66, 2017.

\bibitem{van2008visualizing}
L.~Van~der Maaten and G.~Hinton.
\newblock Visualizing data using t-sne.
\newblock {\em Journal of machine learning research}, 9(11), 2008.

\bibitem{vujivcic2018blockchain}
D.~Vuji{\v{c}}i{\'c}, D.~Jagodi{\'c}, and S.~Ran{\dj}i{\'c}.
\newblock Blockchain technology, bitcoin, and ethereum: A brief overview.
\newblock In {\em 2018 17th international symposium infoteh-jahorina
  (infoteh)}, pages 1--6. IEEE, 2018.

\bibitem{wan2019evaluating}
L.~Wan, D.~Eyers, and H.~Zhang.
\newblock Evaluating the impact of network latency on the safety of blockchain
  transactions.
\newblock In {\em 2019 IEEE International Conference on Blockchain
  (Blockchain)}, pages 194--201. IEEE, 2019.

\bibitem{xiao2020modeling}
Y.~Xiao, N.~Zhang, W.~Lou, and Y.~T. Hou.
\newblock Modeling the impact of network connectivity on consensus security of
  proof-of-work blockchain.
\newblock In {\em IEEE INFOCOM 2020-IEEE Conference on Computer
  Communications}, pages 1648--1657. IEEE, 2020.

\end{thebibliography}

\end{document}